\documentclass[aps,prb,twocolumn,groupedaddress]{revtex4}
\usepackage{graphicx}
\begin{document}

\title{Intrinsic and dislocation induced elastic behavior of solid helium}

\author{James Day}
\altaffiliation{Currently at: Department of Physics and Astronomy, University of British Columbia, Vancouver, British Columbia, Canada V6T 1Z1}
\affiliation{Department of Physics, University of Alberta, Edmonton, Alberta, Canada T6G 2G7}

\author{Oleksandr Syshchenko}
\affiliation{Department of Physics, University of Alberta, Edmonton, Alberta, Canada T6G 2G7}

\author{John Beamish}
\email[]{beamish@phys.ualberta.ca}
\affiliation{Department of Physics, University of Alberta, Edmonton, Alberta, Canada T6G 2G7}

\date{\today}

\begin{abstract}

Recent experiments showed that the shear modulus of solid $^4$He stiffens in the same temperature range (below 200~mK) where mass decoupling and supersolidity have been inferred from torsional oscillator measurements. The two phenomena are clearly related and crystal defects, particularly dislocations, appear to be involved in both. We have studied the effects of annealing and the effects of applying large stresses on the elastic properties of solid $^4$He, using both acoustic resonances and direct low-frequency and low-amplitude measurements of the shear modulus. Both annealing and stressing affect the shear modulus, as expected if dislocations are responsible. However, it is the high temperature modulus which is affected; the low temperature behavior is unchanged and appears to reflect the intrinsic modulus of solid helium. We interpret this behavior in terms of dislocations which are pinned by isotopic $^3$He impurities at low temperatures and so have no effect on the shear modulus. At higher temperatures they become mobile and weaken the solid. Stressing the crystal at low temperatures appears to introduce new defects or additional pinning sites for the dislocation network but these effects can be reversed by heating the crystal above 500~mK. This is in contrast to dislocations produced during crystal growth, which are only annealed at temperatures close to melting.

\end{abstract}


\maketitle

\section{Introduction}

The only measurements which have, thus far, provided direct evidence of the existence of a supersolid state are the torsional oscillator (TO) experiments in which the period decreased at temperatures below about 200~mK~\cite{Kim04-225,Kim04-1921}, suggesting that some fraction of the solid helium within had decoupled from the oscillating probe. This observation has been interpreted in terms of the non-classical rotational inertia (NCRI) which characterizes the supersolid phase. This startling behavior has been replicated by numerous other research  groups~\cite{Kondo07-695,Penzev07-677,Rittner06-165301,Rittner07-175302,Aoki07-015301}, and the observation of a small heat capacity peak supports the existence of a new phase~\cite{Lin07-1025}. On the other hand, several experiments~\cite{Day05-035301,Day06-105304} put upper bounds on possible pressure-driven mass flow which are much smaller than the $\sim$10~$\mu$m/s critical velocities inferred from TO measurements. Other phenomena which would provide unambiguous proof of a superfluid-like state (e.g. persistent currents, quantized vortices, 2$^{nd}$/4$^{th}$ sound) have not yet been observed.

While the fraction of helium which decouples in various TOs varies by three orders of magnitude, the measurements share many features. Decoupling begins around 200~mK, with a gradual onset accompanied by a dissipation peak. It decreases at large oscillation amplitudes, which suggests a superflow critical velocity (v$_c$~$\sim$~10~$\mu$m/s). Decoupling magnitude is independent of oscillation frequency, although its onset shifts with frequency. The behavior depends on the oscillation amplitude during cooling and is hysteretic. The onset temperature is highly sensitive to $^3$He impurities (most measurements used commercial $^4$He gas, with a $^3$He concentration x$_3$~$\sim$~0.3~ppm, but experiments~\cite{Kim07-610,Clark07-135302} with isotopically pure $^4$He, with a $^3$He concentration x$_3$~$\sim$~1~ppb, show a sharper onset at a lower temperature, around 75~mK).

The TO behavior depends on how the solid helium was grown and annealed, indicating that defects are important. Most samples were grown at constant pressure under blocked capillary conditions, a procedure which involves substantial plastic deformation and yields polycrystalline solids with many defects. The magnitude of the decoupling is usually larger in samples grown rapidly or in narrow gaps~\cite{Rittner07-175302} (although the onset temperatures are similar), conditions expected to produce large defect densities. Annealing crystals at high temperatures, which should eliminate some defects, usually reduces the fraction of helium which decouples, sometimes nearly eliminating it~\cite{Rittner06-165301}. The onset of decoupling appears to be sharper in single crystals grown at constant pressure, and in highly annealed crystals~\cite{Clark08-184513}, although these samples still show significant decoupling.  The relevance of sample quality is supported by theoretical work which suggests that supersolidity does not occur in a perfect crystal~\cite{Ceperley04-155303,Prokof'ev05-155302} and that vacancies~\cite{Anderson05-1164}, grain boundaries~\cite{Burovski05-165301,Pollet07-135301}, glassy regions~\cite{Boninsegni06-105301} or dislocations~\cite{Boninsegni07-035301,Toner08-035302} are involved.

The mechanical response of crystals is often determined by dislocations. In this paper, we describe an investigation of the intrinsic and dislocation induced elastic properties of solid helium. We have made direct measurements of the shear modulus of solid $^4$He at low temperatures. This study required the development of a new, although conceptually quite simple, experimental technique. A sample of solid $^4$He is grown in a narrow gap between two parallel plates (piezoelectric shear transducers). One plate, the driving transducer, is moved in a direction parallel to the second plate. The solid $^4$He transmits the resulting elastic shear stress between the plates, and this is measured by the second plate, the detecting transducer. This method allowed us to measure the shear modulus $\mu$ of solid $^4$He directly at strains (stresses) and frequencies significantly lower than previously achievable. We supplement these shear modulus experiments with measurement of the frequency and resonance of acoustic resonances in the surrounding helium.

In an earlier paper~\cite{Day07-853} we showed that the shear modulus $\mu$ of solid $^4$He increases by $\Delta\mu$~$\sim$~10\% at temperatures below 200~mK and that the frequency of an acoustic resonance increases by about 5\% in the same temperature range. We attributed this behavior to dislocations which move in response to external stresses. The modulus anomaly $\Delta\mu$ has the same dependence on temperature, frequency, amplitude and $^3$He concentration as the decoupling observed in TO experiments and the two phenomena are clearly related. In this paper we describe the effects of annealing helium crystals near their melting points and of applying large stresses at low temperatures; these procedures are expected to modify the dislocation network. Both annealing and stressing do affect the shear modulus at high temperatures but they leave the low temperature modulus unchanged. Measurements in hcp $^3$He show the same behavior. These observations are consistent with a dislocation explanation of the modulus changes, since isotopic impurities are expected to bind to dislocations, pinning them and stiffening the solid. The modulus measured at low temperatures, where dislocations are completely immobilized, is the intrinsic value that would be found in defect-free crystals. At high temperatures the impurities unbind, allowing the dislocations to move and reducing the shear modulus.

\section{Experimental design}

In order to compare the elastic properties of solid helium to the behavior seen in torsional oscillator experiments, we developed a new technique to measure the shear modulus of helium at extremely low frequencies and amplitudes. Crystals are described by a set of elastic constants C$_{ij}$ which depend on their structure (e.g. hcp crystals require 5 elastic constants:  C$_{11}$, C$_{12}$, C$_{13}$, C$_{33}$ and C$_{44}$). However, our crystals (and most of those in TO experiments) were grown by the \textquotedblleft blocked capillary" method which is expected to produce polycrystalline samples. These are nearly isotropic and so can be described by two elastic parameters - a bulk modulus K and an effective shear modulus $\mu$. We measured this shear modulus by embedding a pair of piezoelectric shear transducers in solid helium and directly measuring the ratio of stress $\sigma$ to strain $\varepsilon$ in the gap between the transducers. We could measure $\mu$ = $\sigma$/$\varepsilon$ at strains as low as 2.2~x~10$^{-9}$, corresponding to stresses as small as 0.03~Pa. This is two to three orders of magnitude lower than in previous ultrasonic~\cite{Iwasa79-1119,Beamish82-6104,Beamish83-1419}, internal friction~\cite{Tsymbalenko78-787}, and torsional~\cite{Paalanen81-664} experiments and is comparable to the inertial stresses produced in torsional oscillator measurements. A low noise current amplifier allowed us to measure $\mu$ at frequencies down to 20~Hz, far lower than in any other measurements. The small amplitude and low frequency capabilities were indispensable to our measurements. Additionally, our paired piezoelectric shear transducers permitted us to excite and detect acoustic modes outside of the gap separating the transducers and within the surrounding solid $^4$He. The first such acoustic resonance was around 8~kHz and had a quality factor Q~$\sim$~2000 at our lowest temperature. The shear modulus and acoustic resonance measurements in this paper were all made at low drive voltages, in the linear regime where our earlier measurements showed that the behavior was independent of amplitude.

\subsection{Cell construction}

\begin{figure}[htpb]
\begin{center}
\includegraphics[bb=0 0 280 250, scale=0.6]{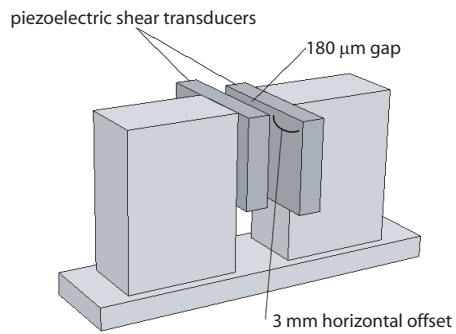}
\caption{Schematic diagram of the parallel-faced piezoelectric shear transducers.} \label{fig:Figure0}
\end{center}
\end{figure}

The shear and acoustic resonance measurements were done in the same experimental cell. Our cell was machined from oxygen-free high-conductivity copper and consisted of a large inner volume ($\approx$~25~cm$^3$) into which our piezoelectric shear transducers were positioned. The transducers were epoxied onto solid brass backing pieces, which were themselves rigidly mounted onto a solid brass support arm, ensuring that the faces between the transducers were parallel, as shown in Figure~\ref{fig:Figure0}. The cell also included an {\it in situ} Straty-Adams capacitive pressure gauge which, when used with a 1~kHz automatic bridge (Andeen-Hagerling~2550A), had a resolution and stability better than 0.2~mbar. The cell was mounted onto the bottom of the mixing chamber of our dilution refrigerator and a 0.004"~i.d. CuNi capillary, thermally anchored at the 1~K pot and the 0.6~K step heat exchangers of the fridge, was used to introduce $^4$He to the cell. The $^4$He used in these experiments ranged in isotopic purity from nominally pure 1~ppb~$^3$He to 0.3~ppm~$^3$He isotopic impurities~\cite{bureau}; the $^3$He crystals in these experiments had a level of 1.35~ppm~$^4$He isotopic impurities. All crystals presented here were grown using the constant volume, blocked capillary technique. Temperatures were measured with a calibrated germanium thermometer, supplemented with a $^{60}$Co nuclear orientation thermometer for temperatures below 50~mK.

\subsection{Measurement}

Displacements were generated and resultant stresses were detected by two parallel-aligned piezoelectric shear transducers~\cite{PZT}, with a helium-filled gap (D~$\approx$~180~$\mu$m) between their faces. The transducers were made from PZT~5A material (quoted fundamental resonance at 500~kHz, with dimensions width W~=~9.6~mm, length L~=~12.8~mm, thickness t~=~2.1~mm). A voltage V applied to the driving transducer produces a shear displacement at its front face $\delta$x~=~d$_{15}$V. These voltages were produced using a synthesized function generator (Stanford Research Systems~DS345), capable of creating many standard waveforms with a frequency resolution of 1~$\mu$Hz. Sinusoidal outputs were used for all measurements, the amplitudes of which were adjustable from 10~mVpp to 10~Vpp. The output signal was split into a driving voltage for the PZT transducer and a reference signal for the 2-phase digital lock-in amplifier (Stanford Research System~SR830~DSP). The minimum amplitude to which the signal generator was set was 150~mVpp, near the reference detection limit of our lock-in amplifier. The driving voltage, however, was further attenuated from 150~mVpp by a series of electronic attenuators and actual driving voltages (used to calculate strains) were measure directly using an auto-ranging microvolt digital multi-meter (Keithley~197).

At room temperature, d$_{15}$ is given by the manufacturer as 5.85~x~10$^{-10}$~m/V. The charge coefficients d$_{ij}$ of a piezo-ceramic are related to its voltage coefficients g$_{ij}$ by the dielectric constant K$_i$ (as in a capacitor the voltage is related to the charge by the capacitance), d$_{ij}$~=~K$_i$$\varepsilon$$_0$g$_{ij}$. In order to determine the low temperature value of the charge coefficient d$_{15}$, consider that the electric field generated in the detecting transducer is E~=~$g_{15}\sigma$; or that the voltage generated across the transducer is V~=~Et~=~$g_{15}\sigma$t, where the stress $\sigma$ is related to the strain $\epsilon$ by the shear modulus ($\mu$~=~$\sigma/\epsilon$) of the helium in the gap D~$\approx$~180~$\mu$m. The strain in the helium in the gap between the two transducers is $\epsilon$~=~$\delta x$/D~=~$d_{15}$V/D, so that the stress induced in the detecting transducer is $\sigma$~=~$\mu d_{15}$V/D.

The capacitance of the piezo-ceramic is given by C~=~K$_1$$\varepsilon_0$A/t, where is A is the area of the electrodes. As q~=~CV, the charge generated on the face of the stressed electrode is q~=~d$_{15}$$\sigma$A. This charge was measured as a generated current I (and at a drive frequency f) I~=~$\omega$q~=~2$\pi$fq~=~2$\pi$fd$_{15}$$\sigma$A. Therefore, the output current from our detecting shear transducer is equal to I~=~2$\pi$fd$_{15}^2$$\mu$VA/D. An ultra-low-noise current preamplifier (Femto~LCA-20K-200M) was used to magnify the signal. The preamplifier has an extremely low 14~fA/$\sqrt{\text{Hz}}$ equivalent input noise current, a 20~kHz bandwidth, and a gain of 2~x~10$^8$~V/A: this signal was then detected with the 2-phase digital lock-in amplifier mentioned above. Upon detection of this signal, d$_{15}$ may be written in terms of our experimental variables at low temperature; in particular, d$_{15}$~=~(DI/2$\pi$A$\mu$Vf)$^{1/2}$.

It is important to note, however, that A is the overlapping area of the electrodes. The total area of each electrode is 1.23~x~10$^{-4}$~m$^2$, but our electrodes are horizontally offset from one another by about 3~mm (refer to Figure~\ref{fig:Figure0}), so that the overlapping area of the transducers was actually 1.0~x~10$^{-4}$~m$^2$. For solid $^4$He at about 35~bar ($\mu$~$\approx$~1.5~x~10$^7$~Pa~\cite{Wilks}), we measured I$_{rms}$~=~25~pA at 2000~Hz for a driving voltage V$_{rms}$~=~17.3~mV. This means that d$_{15}$~=~1.2~x~10$^{-10}$~m/V for these transducers below 4~K (one-fifth of their room temperature value).

So, when a voltage V is applied to the driving transducer, a shear displacement results at its front face $\delta$x~=~d$_{15}$V. Below the resonance frequency of solid $^4$He in the gap (v$_t$/2D $\sim$~830~kHz), this creates a strain, $\epsilon$$_t$~=~$\delta$x/D, which subsequently produces a shear stress, $\sigma$$_t$~=~$\mu$$\epsilon$$_t$, on the detecting transducer. The minimum detectable stress at 2000~Hz, set by noise in our preamplifier (14~fA/$\sqrt{Hz}$, resulting in $\sim$~2.5~fA at 30~s averaging), is $\sigma$$_t$~$\sim$~10$^{-5}$~Pa. This corresponds to a displacement $\delta$x~$\sim$~2~x~10$^{-16}$~m and strain $\epsilon$$_t$~$\sim$~10$^{-12}$. After subtracting a background due to electrical crosstalk (the signal measured with liquid $^4$He in the cell, and a correction of less than 15\%) from the raw signal, the solid's shear modulus ($\sim$~1.5~x~10$^7$~Pa) is proportional to I/f,

\begin{equation}\label{eq:mu}
\mu = \bigg ( \frac{\text{D}}{2 \pi \text{A} \text{d}_{15}^2 \text{V}}\bigg ) \bigg ( \frac{\text{I}}{\text{f}}
\bigg ).
\end{equation}

\noindent This signal has been measured over a frequency range from 20~Hz to 10~kHz, and the shear modulus is essentially frequency independent below 4000~Hz.

While the primary result of the original experiment was the observation of a large anomalous increase in $\mu$ with the same temperature dependence as the decoupling in torsional oscillators, the effect was confirmed by also measuring the frequency f$_r$ and damping 1/Q of an acoustic resonance in the cell. The 3~mm offset of the transducers provided exposed surfaces which could be used to excite and detect acoustic modes of the solid helium outside of the gap, and surrounding the transducers. We expect to observe a resonance in our shear cell around a frequency f$_r$~=~v$_{shear}$/2L, where v$_{shear}$ is the speed of shear sound in the solid and L is the smallest relevant cell dimension. Taking the speed of shear sound in solid $^4$He to be $\sim$~300~m/s and the characteristic length in our shear cell to be $\sim$~2~cm (its diameter), an acoustic resonance in the shear cell was predicted (and observed) near 7500~Hz. To be clear, this is a resonance of the helium in the whole cell and not in the gap between the piezoelectric transducers.

\section{Results}

\begin{figure}[htpb]
\begin{center}
\includegraphics[bb=30 25 675 535, scale=0.38]{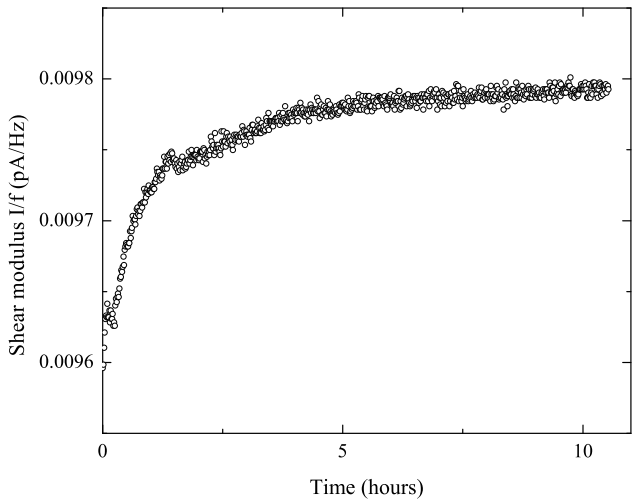}
\caption{Annealing a solid $^4$He sample with 1~ppb~$^3$He impurities, at 33.4~bar and 1.70~K (roughly 200~mK below its melting temperature). Annealing was considered complete when the shear modulus stopped changing with time (which was after about 11 hours for this particular sample).} \label{fig:Graph1}
\end{center}
\end{figure}

Solid samples were grown by first filling the cell with a liquid to high pressure ($\sim$~70~bar) at 4.2~K and then cooling. No special effort was taken to keep the fill line open from the high-pressure gas cylinder to the cell and solidification therefore occurred under constant density conditions. This is commonly referred to as the blocked capillary technique for sample growth, and is expected to produce a polycrystalline solid sample with many defects. One focus of this paper is to probe the effects of annealing on the shear modulus anomaly. Here, the term annealing is meant to describe the process whereby the solid helium sample is kept at a constant temperature slightly below melting for an extended period of time. This process generally improves the quality of a crystal (i.e., results in fewer crystallographic defects). The sample was considered \textquoteleft annealed' when its shear modulus, measured at a constant temperature just below melting, stopped changing as a function of time, as shown in Figure~\ref{fig:Graph1}.

One of the significant features of the torsional oscillator results is the NCRI dependence on sample thermodynamic history. For example, the NCRI fraction can be nearly eliminated through sufficient annealing~\cite{Rittner06-165301}, or can be made remarkably large through rapid quench cooling~\cite{Rittner07-175302} (a process which results in an increased number of crystal defects). Other groups conducting torsional oscillator experiments have noted similar behavior. In the interest of  better understanding the role of defects, we have studied how the shear modulus (and acoustic resonance) of our samples are affected by annealing and by the application of large stresses. These experiments lead to an important conclusion - it is largely the high temperature behavior of the shear modulus that is affected by annealing and stressing the solid, not the low temperature behavior.

\begin{figure}[htpb]
\begin{center}
\includegraphics[bb=35 25 680 540, scale=0.38]{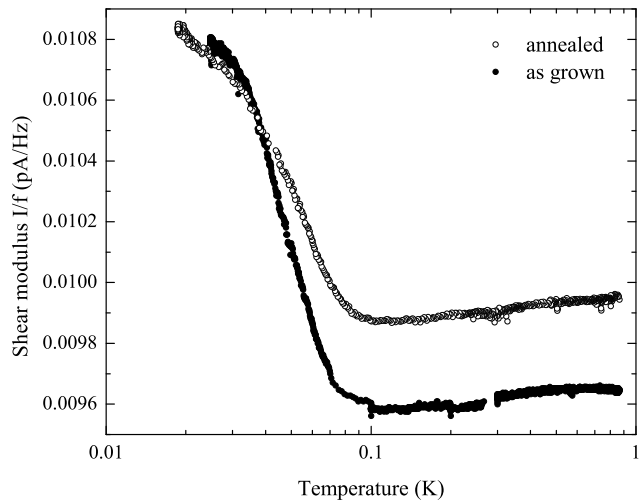}
\caption{Shear modulus of a solid hcp $^4$He sample with 1~ppb~$^3$He impurities,
at 33.4~bar, as a function of temperature, before and after annealing. Both
data sets taken upon cooling.} \label{fig:Graph2}
\end{center}
\end{figure}

Figure~\ref{fig:Graph2} displays the temperature dependence of the shear modulus $\mu$ of the same solid hcp $^4$He sample referenced in Figure~\ref{fig:Graph1}. The measurement was taken at a frequency of 2000~Hz, at a driving voltage of 32.7~mV$_{peak}$, and a corresponding strain of 2.2~x~10$^{-8}$. Note that the annealing process reduced the size of the anomaly $\Delta\mu$~=~($\mu_{\text{low T}}$~-~$\mu_{\text{high T}}$)/$\mu_{\text{low T}}$ from 11.3\% to 7.4\%, but that it was largely the high temperature value of the shear modulus $\mu$ which increased by about 3\%. Put differently, the low temperature value of $\mu$ was essentially unchanged through annealing, i.e. at 25~mK $\mu$ increased by only 0.3\%. Measurements were also taken on several commercial purity $^4$He samples (0.3~ppm~$^3$He) with very similar results: annealing changed the high temperature value of $\mu$ by the order 1\%. In all but one case the high temperature value of $\mu$ increased (it once was reduced by 1\% after annealing).

\begin{figure}[floatfix]
\begin{center}
\includegraphics[bb=40 25 685 535, scale=0.38]{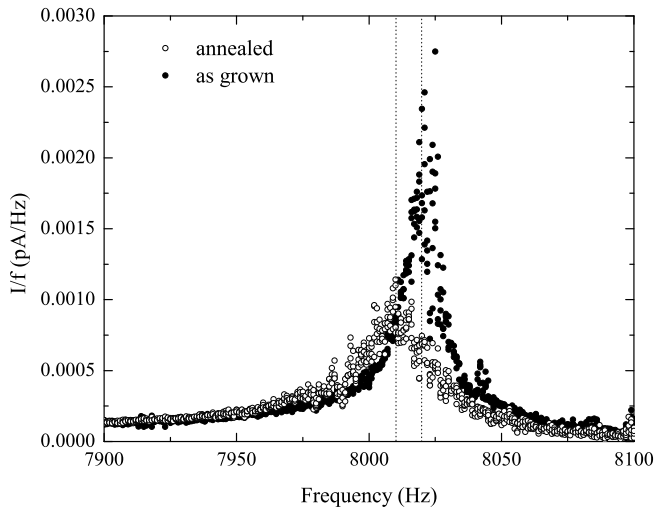}
\caption{Resonance frequency of a solid hcp $^4$He sample with 1~ppb~$^3$He impurities,
at 33.4~bar, at 25~mK, before and after annealing. Dashed lines at resonance center frequencies
are a guide to the eye.} \label{fig:Graph3}
\end{center}
\end{figure}

Similar behavior was also seen in the acoustic resonance. As shown in Figure~\ref{fig:Graph3}, at 25~mK, f$_r$ decreased by only about 0.1\%, from 8020~Hz (as grown) to 8009~Hz (after annealing).

\begin{figure}[htpb]
\begin{center}
\includegraphics[bb=0 15 675 585, scale=0.36]{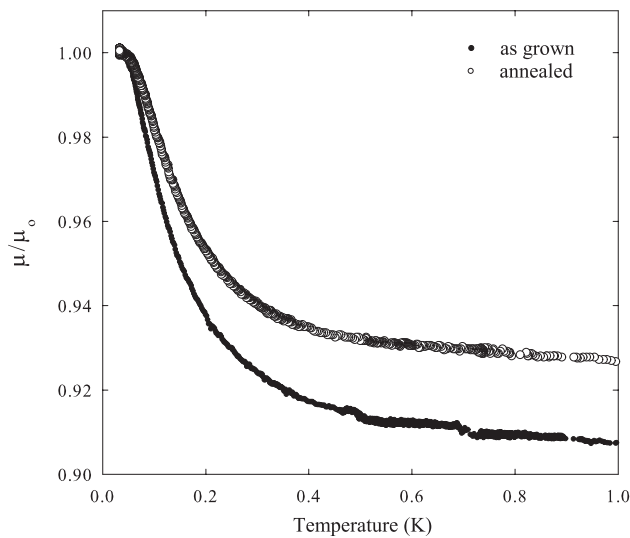}
\caption{Shear modulus of a hcp $^3$He sample (1.35~ppm~$^4$He) at a pressure of 119~bar. Data are taken during cooling, before and after annealing.} \label{fig:Graph4}
\end{center}
\end{figure}

We have also made measurements on hcp $^3$He crystals and observed a similar shear modulus increase at low temperatures. Figure~\ref{fig:Graph4} shows the shear modulus (normalized using the low temperature value for the \textquotedblleft as grown" crystal) for an hcp $^3$He crystal at a pressure of 119~bar. The onset temperature is somewhat higher ($\sim$0.4~K) than in $^4$He, perhaps because of the larger impurity concentration in the $^3$He sample (1.35~ppm~$^4$He, compared to 1~ppb~$^3$He in the $^4$He sample of Figures~\ref{fig:Graph1} to~\ref{fig:Graph3}). Before annealing, the low temperature modulus increase $\Delta\mu$ was about 9\%. Annealing increased the high temperature modulus (by 2\% at 1~K) but barely changed the low temperature value ($\mu$ decreased by 0.02\% at 40~mK), the same behavior seen in hcp $^4$He).

\begin{figure}[htpb]
\begin{center}
\includegraphics[bb=35 30 675 535, scale=0.38]{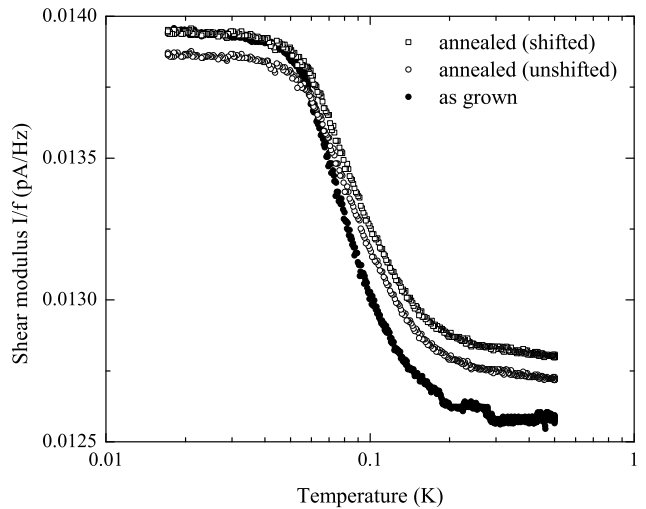}
\caption{Shear modulus of a solid hcp $^4$He sample with 300~ppb~$^3$He impurities,
at 33.3~bar, as a function of temperature, before and after annealing. This sample was annealed
at 1.70~K for 15 hours. Both data sets taken upon cooling. The \textquoteleft annealed'
data is shown as measured (unshifted) and as shifted (to agree at low temperature
with the \textquoteleft as grown' shear modulus). Justification for doing so is
provided in the text.} \label{fig:Graph5}
\end{center}
\end{figure}

\begin{figure*}[htpb]
\begin{center}
\includegraphics[bb=50 25 675 560, scale=0.7]{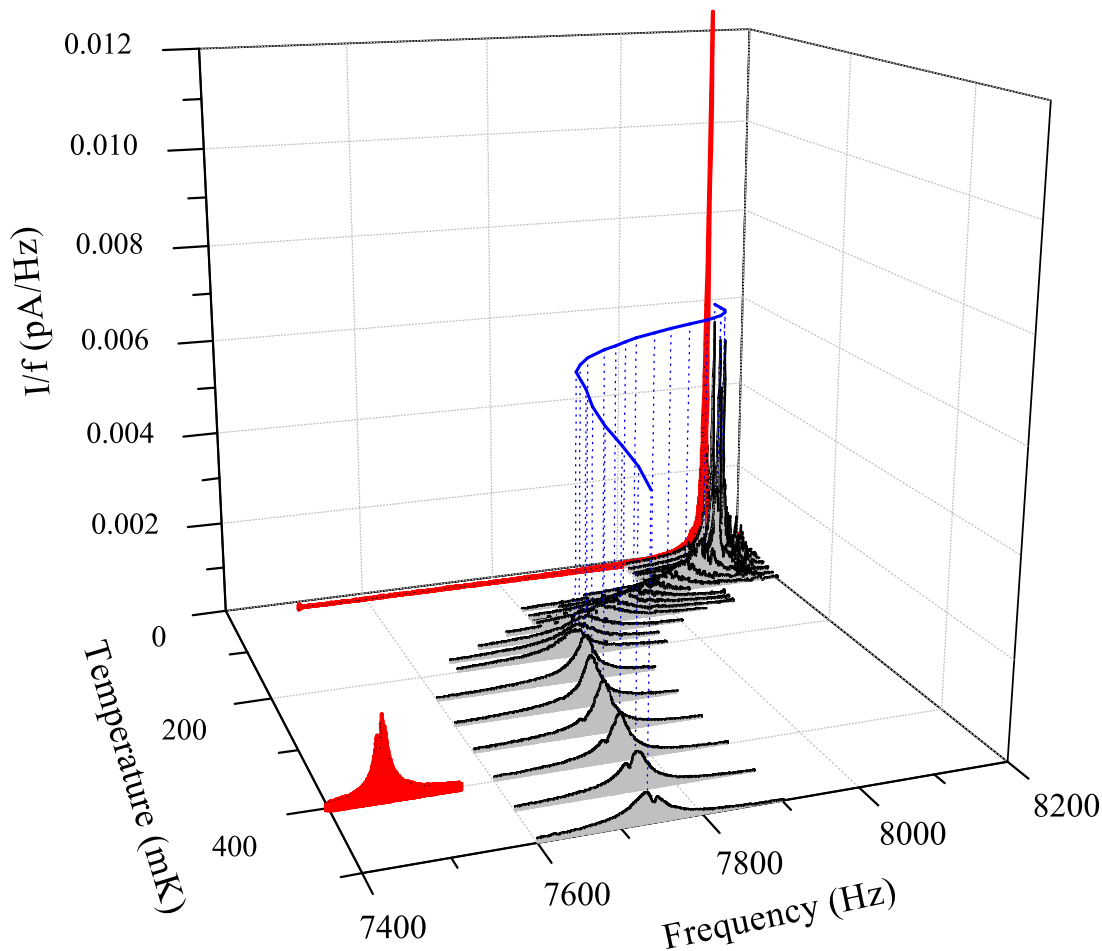}
\caption{Acoustic resonance of a solid hcp $^4$He sample with 300~ppb~$^3$He impurities,
at 33.3~bar, before and after annealing, as a function of temperature. The data in red shows the resonant
peaks at 18~mK and 400~mK before annealing. The blue line traces the path of the resonance through
frequency space as a function of temperature.} \label{fig:Graph6}
\end{center}
\end{figure*}

\begin{figure}[htpb]
\begin{center}
\includegraphics[bb=50 15 680 530, scale=0.38]{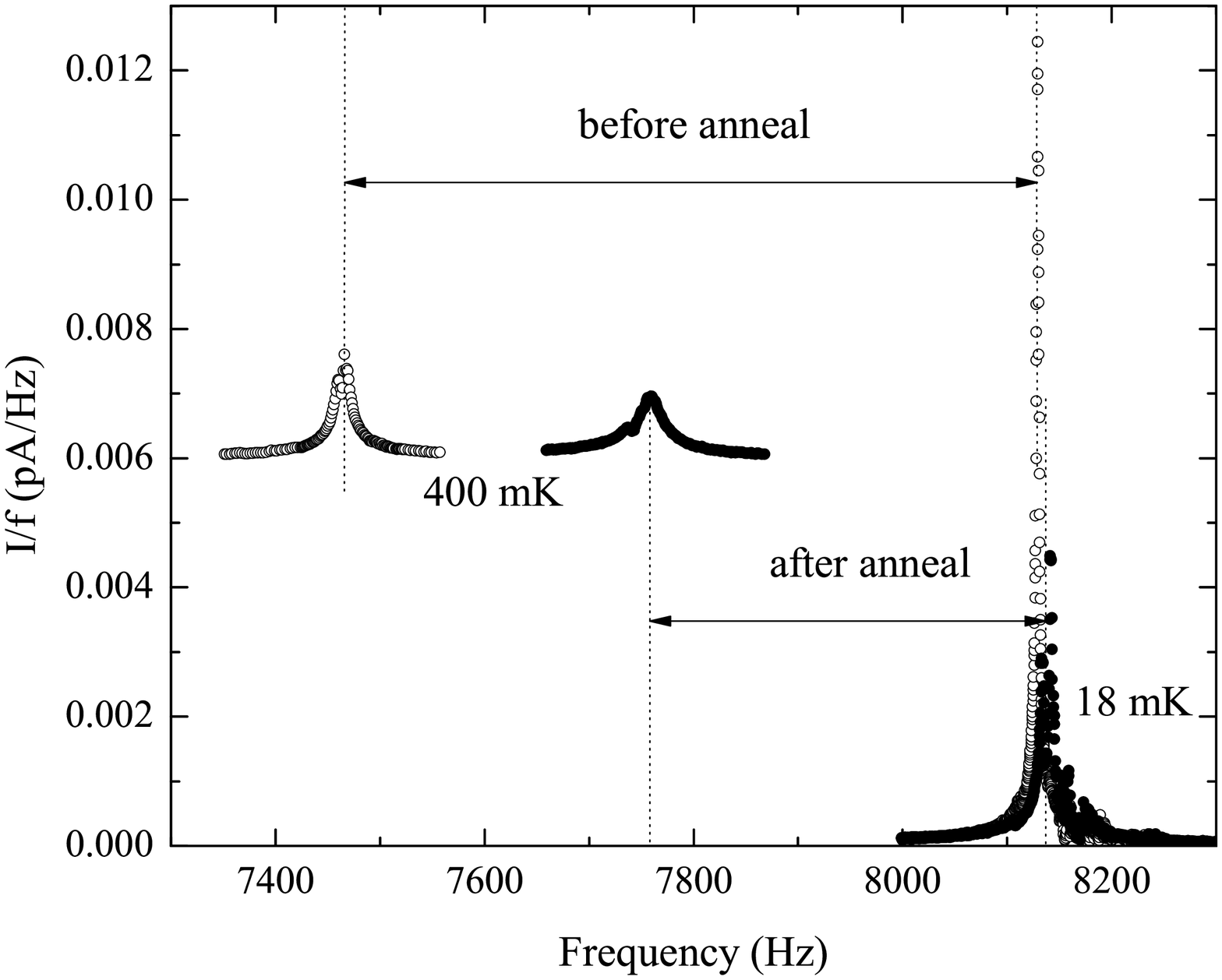}
\caption{Acoustic resonance of the solid hcp $^4$He sample shown in Figure~\ref{fig:Graph5}, before and after annealing, at 18~mK and 400~mK.} \label{fig:Graph12}
\end{center}
\end{figure}

Figures~\ref{fig:Graph5} through~\ref{fig:Graph12} shows the effects of annealing on a second $^4$He sample, this one a 33.3~bar crystal grown from gas with standard isotopic purity ($\sim$~300~ppm~$^3$He). Figures~\ref{fig:Graph6} and~\ref{fig:Graph12}, in particular, show the shear modulus before and after annealing. As in the other samples, annealing reduced $\Delta\mu$, in this case from 9.8 to 7.7\%. The value of $\mu$ decreased by 0.5\% at the lowest temperature (18~mK) but increased by more than 1\% at 400~mK. The acoustic resonance peaks show the effects of annealing even more clearly. Figure~\ref{fig:Graph6} shows the evolution of the peaks as a function of temperature for the annealed sample (see also Figure~3 of reference~\cite{Day07-853}). The peak frequency f$_r$ increases over the same temperature range as the modulus, from 7759~Hz at 400~mK to 8142~Hz at 18~mK. This 5\% increase is about half as large as the corresponding modulus change in Figure~\ref{fig:Graph5}, as is expected since the resonance frequencies are determined by sound speeds which vary as the square root of elastic moduli. The two types of measurements do probe the same elastic behavior, albeit in different parts of the cell (the modulus is measured in the gap between the transducers; the acoustic resonance occurs in the surrounding helium). The red peaks in Figure~\ref{fig:Graph6} are the resonances at 18 and 400~mK before annealing. These are highlighted with their corresponding post-anneal peaks in Figure~\ref{fig:Graph12}. Annealing raised f$_r$ by 4\% at 400~mK but had a very small effect (0.1\% increase) at 18~mK. The resonance frequencies, which depend only on elastic moduli and sample dimensions, are probably more accurate than the modulus values, which can be affected by small changes in amplifier gain or signal crosstalk. Annealing never changes the low temperature resonance frequency by more than ~0.1\%, so we believe that the 0.5\% modulus change in Figure~\ref{fig:Graph5} is not accurate. In Figure~\ref{fig:Graph5} we therefore also show the modulus for the annealed crystal shifted by 0.5\% to agree with the pre-annealing data at low temperature.

Annealing is expected to improve sample quality by, for example, reducing the dislocation density $\Lambda$. Large stresses, on the other hand, can create additional dislocations or can cause existing dislocations to cross, producing new defects like jogs which could act as pinning centers~\cite{Hull}. Large ultrasonic stresses have been shown~\cite{Beamish83-1419} to change the velocity and reduce the attenuation of longitudinal ultrasonic waves in solid helium, effects attributed to pinning of dislocations when they vibrate with large amplitudes. In our experiments, we used the acoustic resonance to produce large stresses by applying large drive voltages at low temperatures (where the resonance's large quality factor, Q~$\sim$~2000, increases the amplitude). This produced stresses of about 700~Pa, much larger than the $\sim$~0.3~Pa level at which our modulus measurements were made.

\begin{figure}[htpb]
\begin{center}
\includegraphics[bb=35 25 675 535, scale=0.38]{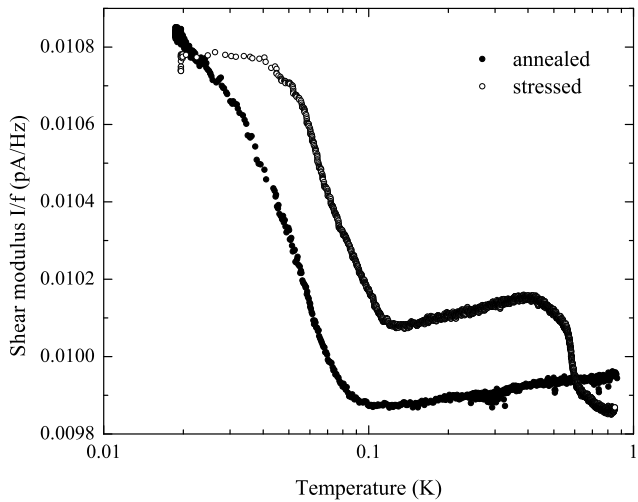}
\caption{Shear modulus of a solid hcp $^4$He sample with 1~ppb~$^3$He impurities,
at 33.4~bar, as a function of temperature, before (upon cooling) and after stressing (upon warming).} \label{fig:Graph9}
\end{center}
\end{figure}

\begin{figure}[htpb]
\begin{center}
\includegraphics[bb=35 25 690 535, scale=0.37]{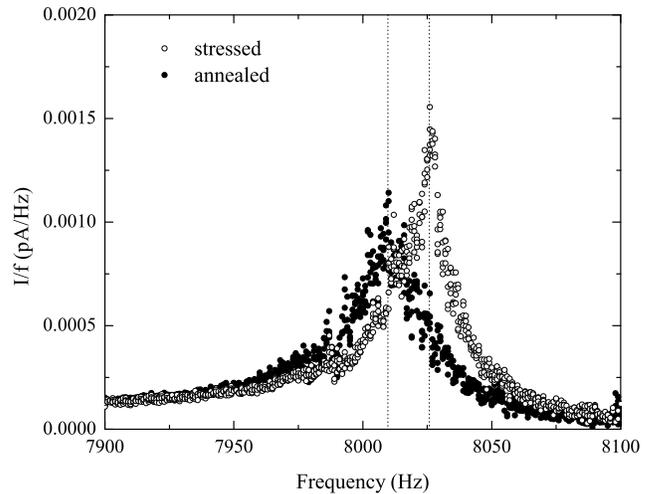}
\caption{Resonance frequency of a solid hcp $^4$He sample with 1~ppb~$^3$He impurities,
at 33.4~bar, at 20~mK, before and after annealing. Dashed lines at resonance center frequencies
are a guide to the eye.} \label{fig:Graph10}
\end{center}
\end{figure}

Figure~\ref{fig:Graph9} shows the effect of stressing the isotopically pure $^4$He crystal. The lower curve is the annealed data from Figure~\ref{fig:Graph2}, while the upper curve is data taken during warming, after stressing the annealed crystal at 19~mK, as described above.  As was the case for annealing, stressing left the value of the low temperature shear modulus essentially unchanged. However, it had a significant effect on the modulus at higher temperature. Upon warming, the modulus remained constant up to about 40~mK, in contrast to the annealed crystal, for which $\mu$ decreased by about 3\% in this temperature range. The modulus then decreased rapidly, giving an onset about 30~mK higher than for the annealed crystal. Also, at high temperature (above 100~mK) $\mu$ was about 2\% larger than in the annealed crystal - stressing thus reduced the modulus anomaly $\Delta\mu$. When this crystal was warmed above 400~mK, the modulus began to decrease (as a function of time) and by 600~mK it had reached its unstressed value. We also looked at how the resonance peaks were affected by stressing the sample. Figure~\ref{fig:Graph10} shows the corresponding low amplitude resonance peaks at 20~mK, before and after applying the large stress. Stressing barely changed the resonance frequency, increasing it by less than 0.2\%.

\begin{figure}[htpb]
\begin{center}
\includegraphics[bb=35 25 675 535, scale=0.38]{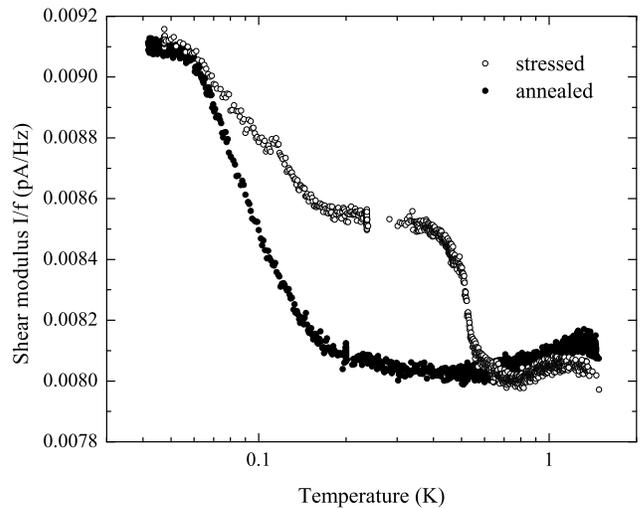}
\caption{Shear modulus of a solid hcp $^4$He sample with 300~ppb~$^3$He impurities,
at 29.3~bar, as a function of temperature, before (upon cooling) and after stressing (upon warming).} \label{fig:Graph11}
\end{center}
\end{figure}

The effects of stressing were essentially the same in samples with higher (300~ppb) $^3$He concentrations. Figure~\ref{fig:Graph11} shows the shear modulus for a 29.3~bar crystal. The lower curve is data taken after annealing for 10 hours at 1.45~K (~$\sim$~0.3~K below melting) and the upper curve was taken during warming, after applying the same stressing procedure at low temperature. Again, the low temperature modulus was unaffected by stressing, the high temperature value (above the onset of the modulus anomaly around 200~mK) increased significantly, and warming above 500~mK reversed the effects. It is worth noting that this was the only crystal for which the effect of the original annealing was to reduce the modulus (and thus increase the modulus anomaly $\Delta\mu$). Annealing also reduced the high temperature resonance frequency by about 2\% but left the low temperature value nearly unchanged (an increase of about 0.1\%). This is in contrast to all other samples, where annealing increased both $\mu$ and f$_r$ at high temperature.

\section{Discussion}

In a previous paper~\cite{Day07-853} we showed that the behavior of the shear modulus in solid $^4$He was remarkably similar to the decoupling seen in TO experiments. The two phenomena had such similar dependences on temperature, frequency, amplitude and isotopic purity that they must be closely related. We interpret the changes in shear modulus in terms of dislocations which move in response to applied stresses and can be pinned by $^3$He impurities. Dislocations in crystals form 3-dimensional networks which are characterized by their Burgers vector b (an interatomic spacing) and density $\Lambda$ (total dislocation length per unit volume). They are pinned where they intersect, at nodes separated by the network loop length L$_N$. Ultrasonic measurements~\cite{Iwasa79-1119,Beamish82-6104,Beamish83-1419} on single crystals of helium give loop lengths in the range L$_N$~=~(5~-10)~x~10$^{-6}$~m. Between pinning points, dislocations can move in their glide plane in response to shear stress. At low frequencies, inertia and phonon damping are unimportant and dislocations simply bow out, like a string under tension. This displacement creates a strain which adds to the elastic strain and reduces the solid's shear modulus. In this regime the modulus change is independent of frequency~\cite{Granato56-583}, $\Delta\mu$/$\mu$~=~-CR$\Lambda$L$^2$, where C is a constant which depends on the distribution of loop lengths ($\sim$~0.1 for a single length, $\sim$~0.5 for an exponential distribution with average length L). The orientation factor R relates the total stress to its component in the glide plane (the basal plane for the dominant slip system of edge dislocations in hcp helium). Long loops have the largest effects on the shear modulus and can dramatically soften the crystal even at low dislocation densities. Fewer dislocations means fewer intersections and therefore longer loops. In annealed crystals with well defined networks,  $\Lambda$L$_N^2$ is a geometric constant (e.g. 3 for a cubic network) so $\Delta\mu$ can be nearly independent of the dislocation density. For the random orientations expected in polycrystalline samples, the average orientation factor is R~$\sim$~0.2, so dislocations can reduce a crystal's intrinsic shear modulus by as much as 30\%, sufficient to explain the ~10\%~-~20\% changes we observe.

The temperature dependence of the modulus is due to the interaction with impurities, which can also pin dislocations, although less strongly than nodes of the network. If the pinning length L$_P$ (the distance between impurity pinning sites) is much smaller than the network length L$_N$, then the dislocations are effectively immobilized and no longer soften the solid, restoring the intrinsic shear modulus of a defect-free crystal.  Impurity pinning is strongly temperature dependent, since impurities are attracted to dislocations with a binding energy E$_B$ (e.g. a $^3$He atom binds to an edge dislocation in hcp $^4$He with energy E$_B$~$\sim$~0.3~-~0.7~K~\cite{Paalanen81-664,Iwasa80-1722}). When a crystal is cooled below E$_B$, impurities condense onto dislocations giving an enhanced concentration x$_D$~=~x$_3$~e$^{\frac{E_B}{k_b T}}$, where x$_3$ is their bulk concentration. Pinning will be significant when x$_D$ increases to the point where a typical dislocation loop has an impurity bound to it (i.e., when x$_D$~$\approx$~a/L$_N$, where a is the atomic spacing along the dislocation). This implies that the shear modulus will recover to its intrinsic value below a pinning temperature which decreases with impurity concentration,

\begin{equation}\label{eq:TP}
\text{T}_P \sim -\frac{\text{E}_B}{\text{k}_B}\frac{1}{ln(\frac{\text{L}_N \text{x}_3}{\text{a}})}.
\end{equation}

Or, more accurately, the shear modulus will soften above this temperature, as impurities unbind from dislocations and allow them to move. The temperature dependence of the modulus changes is consistent with pinning by isotopic impurities. The temperature at which this pinning begins, T$_P$, increases with impurity concentration. Using parameters estimated from ultrasonic measurements~\cite{Iwasa79-1119,Beamish82-6104,Beamish83-1419} on hcp $^4$He (E$_B$~=~0.6~k$_B$, L$_N$~=~5~$\mu$m, a~=~0.35~nm) gives T$_P$~$\sim$~54~mK for the isotopically pure $^4$He sample (1~ppb~$^3$He) of Figure~\ref{fig:Graph2} and T$_P$~$\sim$~110~mK for the commercial $^4$He samples (300~ppb~$^3$He) of Figures~\ref{fig:Graph5} and~\ref{fig:Graph11}. These are close to the onset temperatures for the shear modulus anomaly $\Delta\mu$. The higher onset temperature in the $^3$He crystal of Figure~\ref{fig:Graph4} may reflect its higher impurity concentration (1.35~ppm~$^4$He) but the dislocation parameters may also be different in $^3$He.

In the experiments described above, we have attempted to alter the defect density in two ways: by annealing crystals near melting to eliminate some defects, and by applying large stresses to create new defects. We found that both treatments affect the shear modulus, but only the high temperature behavior. The low temperature modulus is unchanged, as expected since dislocations are completely pinned at temperatures much lower than T$_P$. The modulus measured at low temperatures is the intrinsic modulus of solid helium, i.e. the value which would characterize a defect-free crystal.

The changes in the high temperature modulus produced by annealing and stressing are also consistent with dislocation behavior. The decrease in the shear modulus upon warming, $\Delta\mu$, occurs when impurities no longer pin the dislocations and should be proportional to $\Lambda$L$_N^2$.  Annealing is expected to reduce the dislocation density $\Lambda$ and does usually result in a smaller change $\Delta\mu$. However, reducing $\Lambda$ will increase the network length L$_N$, so that $\Lambda$L$_N^2$ may not be strongly affected by annealing. It can even increase and we did, in one case, observe a larger $\Delta\mu$ after annealing.

Applying large stresses can result in plastic flow and increased dislocation densities~\cite{Hull}. Even at stresses too low for such \textquotedblleft dislocation multiplication", existing dislocations can move across the path of others in the network. This \textquotedblleft forest cutting" mechanism creates new defects on the dislocations (\textquotedblleft jogs") which can act as additional pinning sites. We found that applying large stresses ($\sim$~700~Pa) at low temperatures always increased the high temperature modulus (i.e., reduced the modulus change $\Delta\mu$), which could reflect either dislocation multiplication or jog creation. However, warming the crystals above 500~mK reversed the effect of the applied stress. This is well below the temperature at which we observed significant annealing effects in our original samples and should be contrasted with observations of pressure annealing in flow experiments~\cite{Syshchenko08-427} which, for bulk $^4$He, only happened very close to melting. Jogs are much more easily annealed (by thermal vacancies) and this suggests that, rather than creating new dislocations, the stresses create jogs on existing ones.

The shear modulus anomaly shares many features with the decoupling seen in torsional oscillator measurements~\cite{Day07-853}, evidence that dislocations play an important role in helium's supersolid behavior. Annealing usually reduces both the TO decoupling and the shear modulus softening but can occasionally increase the decoupling~\cite{Penzev07-677} as well as the modulus change. Applying large stresses also reduces the modulus softening, perhaps by creating jogs which can pin dislocations. The response of the shear modulus to annealing and stressing shows that dislocations are completely pinned at the lowest temperatures and only reduce the helium's intrinsic shear modulus when they unpin at high temperature. This suggests that the decoupling in torsional oscillators occurs when dislocations are immobilized and that it may be possible to reduce the decoupling by applying large stresses at low temperatures. Further elastic measurements will help clarify how dislocation properties are related to the sample parameters (e.g. cells' surface to volume ratios or gap thicknesses, crystal growth or quench rates) which appear to be correlated with the TO decoupling.

\begin{acknowledgments}

Funding for this research was provided by the Natural Sciences and Engineering Research Council of Canada and by the University of Alberta. We would like to thank Moses Chan, Harry Kojima, John Reppy, and Keiya Shirahama for valuable discussions.

\end{acknowledgments}

\bibliography{intrinsic}

\begin{thebibliography}{34}
\expandafter\ifx\csname natexlab\endcsname\relax\def\natexlab#1{#1}\fi
\expandafter\ifx\csname bibnamefont\endcsname\relax
  \def\bibnamefont#1{#1}\fi
\expandafter\ifx\csname bibfnamefont\endcsname\relax
  \def\bibfnamefont#1{#1}\fi
\expandafter\ifx\csname citenamefont\endcsname\relax
  \def\citenamefont#1{#1}\fi
\expandafter\ifx\csname url\endcsname\relax
  \def\url#1{\texttt{#1}}\fi
\expandafter\ifx\csname urlprefix\endcsname\relax\def\urlprefix{URL }\fi
\providecommand{\bibinfo}[2]{#2}
\providecommand{\eprint}[2][]{\url{#2}}

\bibitem[{\citenamefont{Kim and Chan}(2004{\natexlab{a}})}]{Kim04-225}
\bibinfo{author}{\bibfnamefont{E.}~\bibnamefont{Kim}} \bibnamefont{and}
  \bibinfo{author}{\bibfnamefont{M.~H.~W.} \bibnamefont{Chan}},
  \bibinfo{journal}{Nature} \textbf{\bibinfo{volume}{427}},
  \bibinfo{pages}{225} (\bibinfo{year}{2004}{\natexlab{a}}).

\bibitem[{\citenamefont{Kim and Chan}(2004{\natexlab{b}})}]{Kim04-1921}
\bibinfo{author}{\bibfnamefont{E.}~\bibnamefont{Kim}} \bibnamefont{and}
  \bibinfo{author}{\bibfnamefont{M.~H.~W.} \bibnamefont{Chan}},
  \bibinfo{journal}{Science} \textbf{\bibinfo{volume}{305}},
  \bibinfo{pages}{1921} (\bibinfo{year}{2004}{\natexlab{b}}).

\bibitem[{\citenamefont{Kondo et~al.}(2007)\citenamefont{Kondo, Takada,
  Shibayama, and Shirahama}}]{Kondo07-695}
\bibinfo{author}{\bibfnamefont{M.}~\bibnamefont{Kondo}},
  \bibinfo{author}{\bibfnamefont{S.}~\bibnamefont{Takada}},
  \bibinfo{author}{\bibfnamefont{Y.}~\bibnamefont{Shibayama}},
  \bibnamefont{and}
  \bibinfo{author}{\bibfnamefont{K.}~\bibnamefont{Shirahama}},
  \bibinfo{journal}{J. Low Temp. Phys.} \textbf{\bibinfo{volume}{148}},
  \bibinfo{pages}{695} (\bibinfo{year}{2007}).

\bibitem[{\citenamefont{Penzev et~al.}(2007)\citenamefont{Penzev, Yasuta, and
  Kubota}}]{Penzev07-677}
\bibinfo{author}{\bibfnamefont{A.}~\bibnamefont{Penzev}},
  \bibinfo{author}{\bibfnamefont{Y.}~\bibnamefont{Yasuta}}, \bibnamefont{and}
  \bibinfo{author}{\bibfnamefont{M.}~\bibnamefont{Kubota}},
  \bibinfo{journal}{J. Low Temp. Phys.} \textbf{\bibinfo{volume}{148}},
  \bibinfo{pages}{677} (\bibinfo{year}{2007}).

\bibitem[{\citenamefont{Rittner and Reppy}(2006)}]{Rittner06-165301}
\bibinfo{author}{\bibfnamefont{A.~S.~C.} \bibnamefont{Rittner}}
  \bibnamefont{and} \bibinfo{author}{\bibfnamefont{J.~D.} \bibnamefont{Reppy}},
  \bibinfo{journal}{Phys. Rev. Lett.} \textbf{\bibinfo{volume}{97}},
  \bibinfo{pages}{165301} (\bibinfo{year}{2006}).

\bibitem[{\citenamefont{Rittner and Reppy}(2007)}]{Rittner07-175302}
\bibinfo{author}{\bibfnamefont{A.~S.~C.} \bibnamefont{Rittner}}
  \bibnamefont{and} \bibinfo{author}{\bibfnamefont{J.~D.} \bibnamefont{Reppy}},
  \bibinfo{journal}{Phys. Rev. Lett.} \textbf{\bibinfo{volume}{98}},
  \bibinfo{pages}{175302} (\bibinfo{year}{2007}).

\bibitem[{\citenamefont{Aoki et~al.}(2007)\citenamefont{Aoki, Graves, and
  Kojima}}]{Aoki07-015301}
\bibinfo{author}{\bibfnamefont{Y.}~\bibnamefont{Aoki}},
  \bibinfo{author}{\bibfnamefont{J.~C.} \bibnamefont{Graves}},
  \bibnamefont{and} \bibinfo{author}{\bibfnamefont{H.}~\bibnamefont{Kojima}},
  \bibinfo{journal}{Phys. Rev. Lett.} \textbf{\bibinfo{volume}{99}},
  \bibinfo{pages}{015301} (\bibinfo{year}{2007}).

\bibitem[{\citenamefont{Lin et~al.}(2007)\citenamefont{Lin, Clark, and
  Chan}}]{Lin07-1025}
\bibinfo{author}{\bibfnamefont{X.}~\bibnamefont{Lin}},
  \bibinfo{author}{\bibfnamefont{A.~C.} \bibnamefont{Clark}}, \bibnamefont{and}
  \bibinfo{author}{\bibfnamefont{M.~H.~W.} \bibnamefont{Chan}},
  \bibinfo{journal}{Nature} \textbf{\bibinfo{volume}{449}},
  \bibinfo{pages}{1025} (\bibinfo{year}{2007}).

\bibitem[{\citenamefont{Day et~al.}(2005)\citenamefont{Day, Herman, and
  Beamish}}]{Day05-035301}
\bibinfo{author}{\bibfnamefont{J.}~\bibnamefont{Day}},
  \bibinfo{author}{\bibfnamefont{T.}~\bibnamefont{Herman}}, \bibnamefont{and}
  \bibinfo{author}{\bibfnamefont{J.}~\bibnamefont{Beamish}},
  \bibinfo{journal}{Phys. Rev. Lett.} \textbf{\bibinfo{volume}{95}},
  \bibinfo{pages}{035301} (\bibinfo{year}{2005}).

\bibitem[{\citenamefont{Day and Beamish}(2006)}]{Day06-105304}
\bibinfo{author}{\bibfnamefont{J.}~\bibnamefont{Day}} \bibnamefont{and}
  \bibinfo{author}{\bibfnamefont{J.}~\bibnamefont{Beamish}},
  \bibinfo{journal}{Phys. Rev. Lett.} \textbf{\bibinfo{volume}{96}},
  \bibinfo{pages}{105304} (\bibinfo{year}{2006}).

\bibitem[{\citenamefont{Kim et~al.}(2007)\citenamefont{Kim, Xia, West, and
  Chan}}]{Kim07-610}
\bibinfo{author}{\bibfnamefont{E.}~\bibnamefont{Kim}},
  \bibinfo{author}{\bibfnamefont{J.~S.} \bibnamefont{Xia}},
  \bibinfo{author}{\bibfnamefont{J.~T.} \bibnamefont{West}}, \bibnamefont{and}
  \bibinfo{author}{\bibfnamefont{M.~H.~W.} \bibnamefont{Chan}},
  \bibinfo{journal}{Bull. Am. Phys. Soc.} \textbf{\bibinfo{volume}{52}},
  \bibinfo{pages}{610} (\bibinfo{year}{2007}).

\bibitem[{\citenamefont{Clark et~al.}(2007)\citenamefont{Clark, West, and
  Chan}}]{Clark07-135302}
\bibinfo{author}{\bibfnamefont{A.~C.} \bibnamefont{Clark}},
  \bibinfo{author}{\bibfnamefont{J.~T.} \bibnamefont{West}}, \bibnamefont{and}
  \bibinfo{author}{\bibfnamefont{M.~H.~W.} \bibnamefont{Chan}},
  \bibinfo{journal}{Phys. Rev. Lett.} \textbf{\bibinfo{volume}{99}},
  \bibinfo{pages}{135302} (\bibinfo{year}{2007}).

\bibitem[{\citenamefont{Clark et~al.}(2008)\citenamefont{Clark, Maynard, and
  Chan}}]{Clark08-184513}
\bibinfo{author}{\bibfnamefont{A.~C.} \bibnamefont{Clark}},
  \bibinfo{author}{\bibfnamefont{J.~D.} \bibnamefont{Maynard}},
  \bibnamefont{and} \bibinfo{author}{\bibfnamefont{M.~H.~W.}
  \bibnamefont{Chan}}, \bibinfo{journal}{Phys. Rev. B}
  \textbf{\bibinfo{volume}{77}}, \bibinfo{pages}{184513}
  (\bibinfo{year}{2008}).

\bibitem[{\citenamefont{Ceperley and Bernu}(2004)}]{Ceperley04-155303}
\bibinfo{author}{\bibfnamefont{D.~M.} \bibnamefont{Ceperley}} \bibnamefont{and}
  \bibinfo{author}{\bibfnamefont{B.}~\bibnamefont{Bernu}},
  \bibinfo{journal}{Phys. Rev. Lett.} \textbf{\bibinfo{volume}{93}},
  \bibinfo{pages}{155303} (\bibinfo{year}{2004}).

\bibitem[{\citenamefont{Prokof'ev and Svistunov}(2005)}]{Prokof'ev05-155302}
\bibinfo{author}{\bibfnamefont{N.}~\bibnamefont{Prokof'ev}} \bibnamefont{and}
  \bibinfo{author}{\bibfnamefont{B.}~\bibnamefont{Svistunov}},
  \bibinfo{journal}{Phys. Rev. Lett.} \textbf{\bibinfo{volume}{94}},
  \bibinfo{pages}{155302} (\bibinfo{year}{2005}).

\bibitem[{\citenamefont{Anderson et~al.}(2005)\citenamefont{Anderson, Brinkman,
  and Huse}}]{Anderson05-1164}
\bibinfo{author}{\bibfnamefont{P.~W.} \bibnamefont{Anderson}},
  \bibinfo{author}{\bibfnamefont{W.~F.} \bibnamefont{Brinkman}},
  \bibnamefont{and} \bibinfo{author}{\bibfnamefont{D.~A.} \bibnamefont{Huse}},
  \bibinfo{journal}{Science} \textbf{\bibinfo{volume}{310}},
  \bibinfo{pages}{1164} (\bibinfo{year}{2005}).

\bibitem[{\citenamefont{Burovski et~al.}(2005)\citenamefont{Burovski, Kozik,
  Kuklov, Prokof'ev, and Svistunov}}]{Burovski05-165301}
\bibinfo{author}{\bibfnamefont{E.}~\bibnamefont{Burovski}},
  \bibinfo{author}{\bibfnamefont{E.}~\bibnamefont{Kozik}},
  \bibinfo{author}{\bibfnamefont{A.}~\bibnamefont{Kuklov}},
  \bibinfo{author}{\bibfnamefont{N.}~\bibnamefont{Prokof'ev}},
  \bibnamefont{and}
  \bibinfo{author}{\bibfnamefont{B.}~\bibnamefont{Svistunov}},
  \bibinfo{journal}{Phys. Rev. Lett.} \textbf{\bibinfo{volume}{94}},
  \bibinfo{pages}{165301} (\bibinfo{year}{2005}).

\bibitem[{\citenamefont{Pollet et~al.}(2007)\citenamefont{Pollet, Boninsegni,
  Kuklov, V., Svistunov, and Troyer}}]{Pollet07-135301}
\bibinfo{author}{\bibfnamefont{L.}~\bibnamefont{Pollet}},
  \bibinfo{author}{\bibfnamefont{M.}~\bibnamefont{Boninsegni}},
  \bibinfo{author}{\bibfnamefont{A.~B.} \bibnamefont{Kuklov}},
  \bibinfo{author}{\bibfnamefont{P.~N.} \bibnamefont{V.}},
  \bibinfo{author}{\bibfnamefont{B.~V.} \bibnamefont{Svistunov}},
  \bibnamefont{and} \bibinfo{author}{\bibfnamefont{M.}~\bibnamefont{Troyer}},
  \bibinfo{journal}{Phys. Rev. Lett.} \textbf{\bibinfo{volume}{98}},
  \bibinfo{pages}{135301} (\bibinfo{year}{2007}).

\bibitem[{\citenamefont{Boninsegni et~al.}(2006)\citenamefont{Boninsegni,
  Prokof'ev, and Svistunov}}]{Boninsegni06-105301}
\bibinfo{author}{\bibfnamefont{M.}~\bibnamefont{Boninsegni}},
  \bibinfo{author}{\bibfnamefont{N.}~\bibnamefont{Prokof'ev}},
  \bibnamefont{and}
  \bibinfo{author}{\bibfnamefont{B.}~\bibnamefont{Svistunov}},
  \bibinfo{journal}{Phys. Rev. Lett.} \textbf{\bibinfo{volume}{96}},
  \bibinfo{pages}{105301} (\bibinfo{year}{2006}).

\bibitem[{\citenamefont{Boninsegni et~al.}(2007)\citenamefont{Boninsegni,
  Kuklov, Pollet, Prokof'ev, Svistunov, and Troyer}}]{Boninsegni07-035301}
\bibinfo{author}{\bibfnamefont{M.}~\bibnamefont{Boninsegni}},
  \bibinfo{author}{\bibfnamefont{A.~B.} \bibnamefont{Kuklov}},
  \bibinfo{author}{\bibfnamefont{L.}~\bibnamefont{Pollet}},
  \bibinfo{author}{\bibfnamefont{N.~V.} \bibnamefont{Prokof'ev}},
  \bibinfo{author}{\bibfnamefont{B.~V.} \bibnamefont{Svistunov}},
  \bibnamefont{and} \bibinfo{author}{\bibfnamefont{M.}~\bibnamefont{Troyer}},
  \bibinfo{journal}{Phys. Rev. Lett.} \textbf{\bibinfo{volume}{99}},
  \bibinfo{pages}{035301} (\bibinfo{year}{2007}).

\bibitem[{\citenamefont{Toner}(2008)}]{Toner08-035302}
\bibinfo{author}{\bibfnamefont{J.}~\bibnamefont{Toner}},
  \bibinfo{journal}{Phys. Rev. Lett.} \textbf{\bibinfo{volume}{100}},
  \bibinfo{pages}{035302} (\bibinfo{year}{2008}).

\bibitem[{\citenamefont{Day and Beamish}(2007)}]{Day07-853}
\bibinfo{author}{\bibfnamefont{J.}~\bibnamefont{Day}} \bibnamefont{and}
  \bibinfo{author}{\bibfnamefont{J.}~\bibnamefont{Beamish}},
  \bibinfo{journal}{Nature} \textbf{\bibinfo{volume}{450}},
  \bibinfo{pages}{853} (\bibinfo{year}{2007}).

\bibitem[{\citenamefont{Iwasa et~al.}(1979)\citenamefont{Iwasa, Araki, and
  Suzuki}}]{Iwasa79-1119}
\bibinfo{author}{\bibfnamefont{I.}~\bibnamefont{Iwasa}},
  \bibinfo{author}{\bibfnamefont{K.}~\bibnamefont{Araki}}, \bibnamefont{and}
  \bibinfo{author}{\bibfnamefont{H.}~\bibnamefont{Suzuki}},
  \bibinfo{journal}{J. Phys. Soc. Japan} \textbf{\bibinfo{volume}{46}},
  \bibinfo{pages}{1119} (\bibinfo{year}{1979}).

\bibitem[{\citenamefont{Beamish and Franck}(1982)}]{Beamish82-6104}
\bibinfo{author}{\bibfnamefont{J.~R.} \bibnamefont{Beamish}} \bibnamefont{and}
  \bibinfo{author}{\bibfnamefont{J.~P.} \bibnamefont{Franck}},
  \bibinfo{journal}{Phys. Rev. B} \textbf{\bibinfo{volume}{26}},
  \bibinfo{pages}{6104} (\bibinfo{year}{1982}).

\bibitem[{\citenamefont{Beamish and Franck}(1983)}]{Beamish83-1419}
\bibinfo{author}{\bibfnamefont{J.~R.} \bibnamefont{Beamish}} \bibnamefont{and}
  \bibinfo{author}{\bibfnamefont{J.~P.} \bibnamefont{Franck}},
  \bibinfo{journal}{Phys. Rev. B} \textbf{\bibinfo{volume}{28}},
  \bibinfo{pages}{1419} (\bibinfo{year}{1983}).

\bibitem[{\citenamefont{Tsymbalenko}(1978)}]{Tsymbalenko78-787}
\bibinfo{author}{\bibfnamefont{V.~L.} \bibnamefont{Tsymbalenko}},
  \bibinfo{journal}{Sov. Phys. JETP} \textbf{\bibinfo{volume}{47}},
  \bibinfo{pages}{787} (\bibinfo{year}{1978}).

\bibitem[{\citenamefont{Paalanen et~al.}(1981)\citenamefont{Paalanen, Bishop,
  and Dail}}]{Paalanen81-664}
\bibinfo{author}{\bibfnamefont{M.~A.} \bibnamefont{Paalanen}},
  \bibinfo{author}{\bibfnamefont{D.~J.} \bibnamefont{Bishop}},
  \bibnamefont{and} \bibinfo{author}{\bibfnamefont{H.~W.} \bibnamefont{Dail}},
  \bibinfo{journal}{Phys. Rev. Lett.} \textbf{\bibinfo{volume}{46}},
  \bibinfo{pages}{664} (\bibinfo{year}{1981}).

\bibitem[{bur()}]{bureau}
\emph{\bibinfo{title}{obtained from {U}.{S}. {B}ureau of {M}ines, {A}marillo,
  {T}exas.}}

\bibitem[{PZT()}]{PZT}
\emph{\bibinfo{title}{www.bostonpiezooptics.com}}.

\bibitem[{\citenamefont{Wilks}(1967)}]{Wilks}
\bibinfo{author}{\bibfnamefont{J.}~\bibnamefont{Wilks}},
  \emph{\bibinfo{title}{The Properties of Liquid and Solid Helium}}
  (\bibinfo{publisher}{Clarendon Press}, \bibinfo{address}{Oxford},
  \bibinfo{year}{1967}).

\bibitem[{\citenamefont{Hull and Bacon}(2001)}]{Hull}
\bibinfo{author}{\bibfnamefont{D.}~\bibnamefont{Hull}} \bibnamefont{and}
  \bibinfo{author}{\bibfnamefont{D.~J.} \bibnamefont{Bacon}},
  \emph{\bibinfo{title}{Introduction to Dislocations}}
  (\bibinfo{publisher}{Butterworth-Heinemann}, \bibinfo{address}{Oxford},
  \bibinfo{year}{2001}).

\bibitem[{\citenamefont{Granato and Lucke}(1956)}]{Granato56-583}
\bibinfo{author}{\bibfnamefont{A.}~\bibnamefont{Granato}} \bibnamefont{and}
  \bibinfo{author}{\bibfnamefont{K.}~\bibnamefont{Lucke}}, \bibinfo{journal}{J.
  Appl. Phys.} \textbf{\bibinfo{volume}{27}}, \bibinfo{pages}{583}
  (\bibinfo{year}{1956}).

\bibitem[{\citenamefont{Iwasa and Suzuki}(1980)}]{Iwasa80-1722}
\bibinfo{author}{\bibfnamefont{I.}~\bibnamefont{Iwasa}} \bibnamefont{and}
  \bibinfo{author}{\bibfnamefont{H.}~\bibnamefont{Suzuki}},
  \bibinfo{journal}{J. Phys. Soc. Japan} \textbf{\bibinfo{volume}{49}},
  \bibinfo{pages}{1722} (\bibinfo{year}{1980}).

\bibitem[{\citenamefont{Syshchenko et~al.}(2008)\citenamefont{Syshchenko, Day,
  and Beamish}}]{Syshchenko08-427}
\bibinfo{author}{\bibfnamefont{A.}~\bibnamefont{Syshchenko}},
  \bibinfo{author}{\bibfnamefont{J.}~\bibnamefont{Day}}, \bibnamefont{and}
  \bibinfo{author}{\bibfnamefont{J.}~\bibnamefont{Beamish}},
  \bibinfo{journal}{Fiz. Nizk. Temp.} \textbf{\bibinfo{volume}{34}},
  \bibinfo{pages}{427} (\bibinfo{year}{2008}).

\end{thebibliography}

\end{document}